\documentclass[%
 reprint,
superscriptaddress,
 amsmath,amssymb,
 aps,
]{revtex4-2}
\usepackage[colorlinks,bookmarks=false,citecolor=magenta,linkcolor=magenta,urlcolor=
magenta]{hyperref}
\usepackage[normalem]{ulem}  
\usepackage{amsmath,amssymb}
\usepackage{setspace}
\usepackage{graphicx}
\usepackage{float}
\usepackage{dcolumn}
\usepackage{bm}
\usepackage{subfigure}
\usepackage{pifont} 
\usepackage{braket}
\usepackage{siunitx}
\usepackage{booktabs}
\usepackage{array}
\usepackage{natbib}
\usepackage{multirow}
\usepackage{xcolor}
\usepackage{natbib}
\usepackage{makecell}
\setcitestyle{numbers,square}

\usepackage[normalem]{ulem} 

\definecolor{mypink}{rgb}{1.0, 0.3, 0.6}

\usepackage{changes}

\begin{document}

\preprint{APS/123-QED}

\title{Stable and efficient charging of superconducting capacitively shunted flux quantum batteries}

\author{Li Li}
\thanks{These authors contributed equally to this work.}
\affiliation{
Beijing National Laboratory for Condensed Matter Physics, 
Institute of Physics, Chinese Academy of Sciences, Beijing 100190, China
}
\affiliation{
School of Physical Sciences, University of Chinese Academy of Sciences, Beijing 100049, China
}
\author{Si-Lu Zhao}
\thanks{These authors contributed equally to this work.}
\affiliation{
Beijing National Laboratory for Condensed Matter Physics, 
Institute of Physics, Chinese Academy of Sciences, Beijing 100190, China
}
\affiliation{
School of Physical Sciences, University of Chinese Academy of Sciences, Beijing 100049, China
}

\author{Yun-Hao Shi}
\affiliation{
Beijing National Laboratory for Condensed Matter Physics, 
Institute of Physics, Chinese Academy of Sciences, Beijing 100190, China
}

\author{Bing-Jie Chen}
\affiliation{
Beijing National Laboratory for Condensed Matter Physics, 
Institute of Physics, Chinese Academy of Sciences, Beijing 100190, China
}
\affiliation{
School of Physical Sciences, University of Chinese Academy of Sciences, Beijing 100049, China
}

\author{Xinhui Ruan}
\affiliation{
Beijing National Laboratory for Condensed Matter Physics, 
Institute of Physics, Chinese Academy of Sciences, Beijing 100190, China
}
\affiliation{
Department of Automation, Tsinghua University, Beijing 100084, P. R. China
}
\author{Gui-Han Liang}
\affiliation{
Beijing National Laboratory for Condensed Matter Physics, 
Institute of Physics, Chinese Academy of Sciences, Beijing 100190, China
}
\affiliation{
School of Physical Sciences, University of Chinese Academy of Sciences, Beijing 100049, China
}

\author{Wei-Ping Yuan}
\affiliation{
Beijing National Laboratory for Condensed Matter Physics, 
Institute of Physics, Chinese Academy of Sciences, Beijing 100190, China
}
\affiliation{
School of Physical Sciences, University of Chinese Academy of Sciences, Beijing 100049, China
}
\author{Jia-Cheng Song}
\affiliation{
Beijing National Laboratory for Condensed Matter Physics, 
Institute of Physics, Chinese Academy of Sciences, Beijing 100190, China
}
\affiliation{
School of Physical Sciences, University of Chinese Academy of Sciences, Beijing 100049, China
}

\author{Cheng-Lin Deng}
\affiliation{
Beijing National Laboratory for Condensed Matter Physics, 
Institute of Physics, Chinese Academy of Sciences, Beijing 100190, China
}
\affiliation{
School of Physical Sciences, University of Chinese Academy of Sciences, Beijing 100049, China
}

\author{Yu Liu}
\affiliation{
Beijing National Laboratory for Condensed Matter Physics, 
Institute of Physics, Chinese Academy of Sciences, Beijing 100190, China
}
\affiliation{
School of Physical Sciences, University of Chinese Academy of Sciences, Beijing 100049, China
}

\author{Tian-Ming Li}
\affiliation{
Beijing National Laboratory for Condensed Matter Physics, 
Institute of Physics, Chinese Academy of Sciences, Beijing 100190, China
}
\affiliation{
School of Physical Sciences, University of Chinese Academy of Sciences, Beijing 100049, China
}

\author{Zheng-He Liu}
\affiliation{
Beijing National Laboratory for Condensed Matter Physics, 
Institute of Physics, Chinese Academy of Sciences, Beijing 100190, China
}
\affiliation{
School of Physical Sciences, University of Chinese Academy of Sciences, Beijing 100049, China
}

\author{Xue-Yi Guo}
\affiliation{
Beijing Academy of Quantum Information Sciences, Beijing 100193, China
}

\author{Xiaohui Song}
\affiliation{
Beijing National Laboratory for Condensed Matter Physics, 
Institute of Physics, Chinese Academy of Sciences, Beijing 100190, China
}
\affiliation{
School of Physical Sciences, University of Chinese Academy of Sciences, Beijing 100049, China
}

\author{Kai Xu}
\email{kaixu@iphy.ac.cn}

\affiliation{
Beijing National Laboratory for Condensed Matter Physics, 
Institute of Physics, Chinese Academy of Sciences, Beijing 100190, China
}
\affiliation{
School of Physical Sciences, University of Chinese Academy of Sciences, Beijing 100049, China
}
\affiliation{
Beijing Academy of Quantum Information Sciences, Beijing 100193, China
}
\affiliation{
Hefei National Laboratory, Hefei 230088, China
}

\author{Heng Fan}
\email{hfan@iphy.ac.cn}

\affiliation{
Beijing National Laboratory for Condensed Matter Physics, 
Institute of Physics, Chinese Academy of Sciences, Beijing 100190, China
}
\affiliation{
School of Physical Sciences, University of Chinese Academy of Sciences, Beijing 100049, China
}
\affiliation{
Beijing Academy of Quantum Information Sciences, Beijing 100193, China
}
\affiliation{
Hefei National Laboratory, Hefei 230088, China
}

\author{Zhongcheng Xiang}
\email{zcxiang@iphy.ac.cn}

\affiliation{
Beijing National Laboratory for Condensed Matter Physics, 
Institute of Physics, Chinese Academy of Sciences, Beijing 100190, China
}
\affiliation{
School of Physical Sciences, University of Chinese Academy of Sciences, Beijing 100049, China
}
\affiliation{
Hefei National Laboratory, Hefei 230088, China
}

\author{Dongning Zheng}
\email{dzheng@iphy.ac.cn}
\affiliation{
Beijing National Laboratory for Condensed Matter Physics, 
Institute of Physics, Chinese Academy of Sciences, Beijing 100190, China
}
\affiliation{
School of Physical Sciences, University of Chinese Academy of Sciences, Beijing 100049, China
}
\affiliation{
Hefei National Laboratory, Hefei 230088, China
}

\begin{abstract}
Quantum batteries, as miniature energy storage devices, have sparked significant research interest in recent years. However, achieving rapid and stable energy transfer in quantum batteries while obeying quantum speed limits remains a critical challenge. In this work, we experimentally optimize the charging process by leveraging the unique energy level structure of a superconducting capacitively-shunted flux qubit, using counterdiabatic pulses in the stimulated Raman adiabatic passage. Compared to previous studies, we impose two different norm constraints on the driving Hamiltonian, achieving optimal charging without exceeding the overall driving strength. Furthermore, we experimentally demonstrate a charging process that achieves the quantum speed limit. In addition, we introduce a dimensionless parameter $\mathcal{S}$ to unify charging speed and stability, offering a universal metric for performance optimization. In contrast to metrics such as charging power and thermodynamic efficiency, the $\mathcal{S}$ criterion quantitatively captures the stability of ergentropy while also considering the charging speed. Our results highlight the potential of the capacitively-shunted qubit platform as an ideal candidate for realizing three-level quantum batteries and deliver novel strategies for optimizing energy transfer protocols.
\end{abstract}
\maketitle
\section{Introduction\label{sec1}}
\begin{figure}
    \centering
    \includegraphics[height = 6.8 cm]{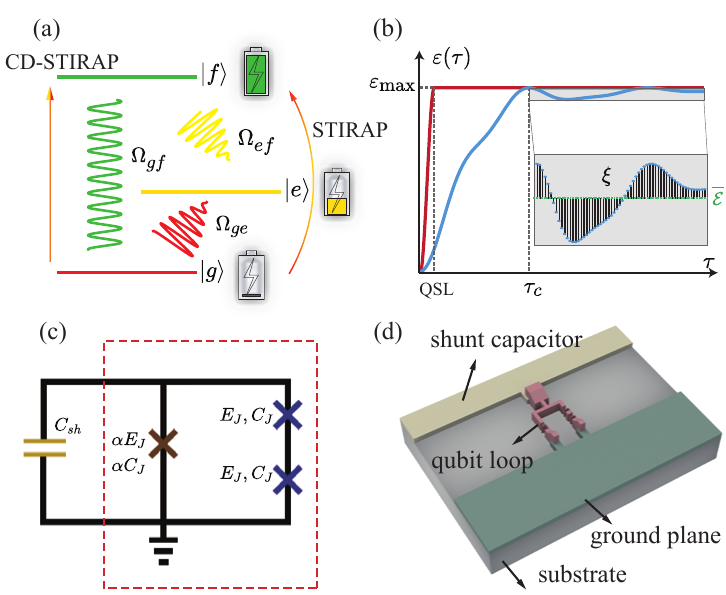}
    \caption{Schematic representation of the charging method and the C-shunt flux qubit device. (a)~Illustration of the STIRAP and CD-STIRAP methods in a three-level system. (b)~Steady-state charging curve of the ergotropy, where $\tau_c$ denotes the charging time when the maximum value is first reached. The standard deviation of the ergotropy for $\tau > \tau_c$ is denoted as $\xi$, which is used to assess the stability of the charging process after it completes. The enlarged grey region corresponds to the part of $\mathcal{E}$ with $\tau > \tau_c$. The green dashed line indicates the average value $\overline{\mathcal{E}}$ in this region. The maximum ergotropy is denote as $\mathcal{E}_\text{max}$, which is equal to the energy of the highest energy level in the quantum system. The blue curve is obtained from numerical simulations based on the STIRAP method, while the red curve corresponds to simulations based on the quantum speed limit (QSL). (c)~Schematic diagram of the C-shunt flux qubit. (d)~3D schematic representation of the C-shunt flux qubit, which shows the actual structure of the qubit.}
    \label{fig1}
\end{figure}

Recent advances in quantum engineering have spurred significant progress in the development of miniaturized devices~\cite{hodaei2014,osullivan2022,song2024b,tobar2024}. Among these developments, quantum batteries~(QBs) have attracted considerable attention as promising candidates for next-generation microenergy storage systems, particularly for their potential applications in quantum computing~\cite{chiribella2021,auffeves2022}, quantum metrology~\cite{chiribella2017,albarelli2020} and other practical tasks~\cite{altman2021a,atzori2019,deutsch2020,acin2018}. For example, QBs can provide precisely controllable energy support for reversible quantum computation, enabling quantum gate operations to be performed efficiently while conserving the total energy, and allowing the QB state to be recycled across multiple computational steps, thus achieving energy-efficient quantum processing~\cite{chiribella2021}. Beyond their technological promise, QBs also provide a unique platform for exploring fundamental quantum thermodynamic principles, exploiting intrinsic quantum resources such as superposition and entanglement~\cite{campaioli2024,hovhannisyan2013,skrzypczyk2014,bruschi2015,korzekwa2016,francica2020,gyhm2024}. These quantum properties enable QBs to achieve charging speeds that surpass those of classical energy storage devices~\cite{ferraro2018,francica2019,vonlindenfels2019,cakmak2020}, as quantified by ergotropy and power. Ergotropy is the maximum work extractable through cyclic processes, while power is the ergotropy change within a given time interval~\cite{allahverdyan2004,alicki2013,kamin2020,shi2022,liu2021,huang2023,monsel2020,garcia-pintos2020,andolina2019,zhao2021a,niu2024}. In addition to the quantum advantages of many-body QBs~\cite{dou2022,mojaveri2023,joshi2022,rossini2020}, research on the charging dynamics of single-body QB is also a key topic of current interest~\cite{gemme2022,yang2024c,yang2024e,hu2022,elghaayda2025}. Among the proposed implementations, three-level QBs~\cite{yang2024e,yang2024c,dou2022b,gemme2024} utilizing Stimulated Raman Adiabatic Passage (STIRAP) have gained prominence for their stable charging characteristics~\cite{santos2019}. Recently, both theoretical~\cite{dou2022a,dou2020} and experimental~\cite{ge2023} progress was made on schemes using counter-diabatic~(CD) driving method, a technique within the shortcut-to-adiabaticity (STA) framework~\cite{guery-odelin2019,wu2024,schaff2011,zhang2013a,an2016,du2016,zhou2017,kolbl2019}, to further improve charging speeds. The underlying principle of STIRAP and CD-STIRAP method is schematically illustrated in Fig.~\ref{fig1}(a).

However, current optimization strategies for single-body three-level QB face two fundamental limitations. First, they disregard constraints on total external field strength, particularly when CD pulses increase the driving Hamiltonian norm~\cite{campaioli2017,binder2015}, creating inequitable comparisons with standard STIRAP~\cite{dou2022a,dou2020,ge2023}. Second, conventional power metrics fail to adequately characterize practical charging processes, as they do not capture the persistent ergotropy oscillations that emerge after reaching the initial maximum. These unwanted oscillations, which may result from imperfect adiabatic approximations during rapid charging or experimental pulse distortion, undermine the stability of QB charging~\cite{hu2022}.

In this work, we systematically optimize CD-STIRAP protocols under a total Hamiltonian norm constraint applied to a superconducting capacitively-shunted~(C-shunt) flux qubit~\cite{yan2016}, featuring accessible $|g\rangle \leftrightarrow |f\rangle$ transitions~\cite{liu2005a} and large positive anharmonicity~\cite{yan2016,yan2020,abdurakhimov2019}. To comprehensively evaluate the charging method, we establish a comprehensive charging evaluation framework as illustrated in Fig.~\ref{fig1}(b). We introduce two key parameters in QBs. The first, $\tau_c$, is defined as the time at which the ergotropy first reaches a local maximum. The second, $\xi = \sqrt{\frac{1}{N} \sum_{i=1}^{N} \big(\mathcal{E}(\tau_c + N\Delta \tau)-\overline{\mathcal{E}}\big)^2}$ with $\overline{\mathcal{E}} = \frac{1}{N} \sum_{i=1}^{N} \mathcal{E}(\tau_c + N\Delta \tau)$, is the standard deviation of the ergotropy after it reaches the first local maximum. It quantifies the stability of the charging process. These parameters are combined into a dimensionless figure of merit, $\mathcal{S} = 1/(\tau_c \xi)$, where large $\mathcal{S}$ indicates faster charging with enhanced stability. Under this metric, we establish the charging advantage of the CD-STIRAP technique over the STIRAP technique, while adhering to the constraint on the total driving strength, highlighting the significance of the $\mathcal{S}$ metric. Furthermore, we establish the CD-STIRAP approach as an effective charging protocol even under different constraints, addressing a critical condition that was previously overlooked in the literature. Finally, we experimentally realize the quantum speed limit~(QSL) charging process, showcasing the C-shunt flux qubit as a promising platform for the development of three-level QBs.

\begin{figure*}[t]
    \centering
    \includegraphics[height = 9.4 cm]{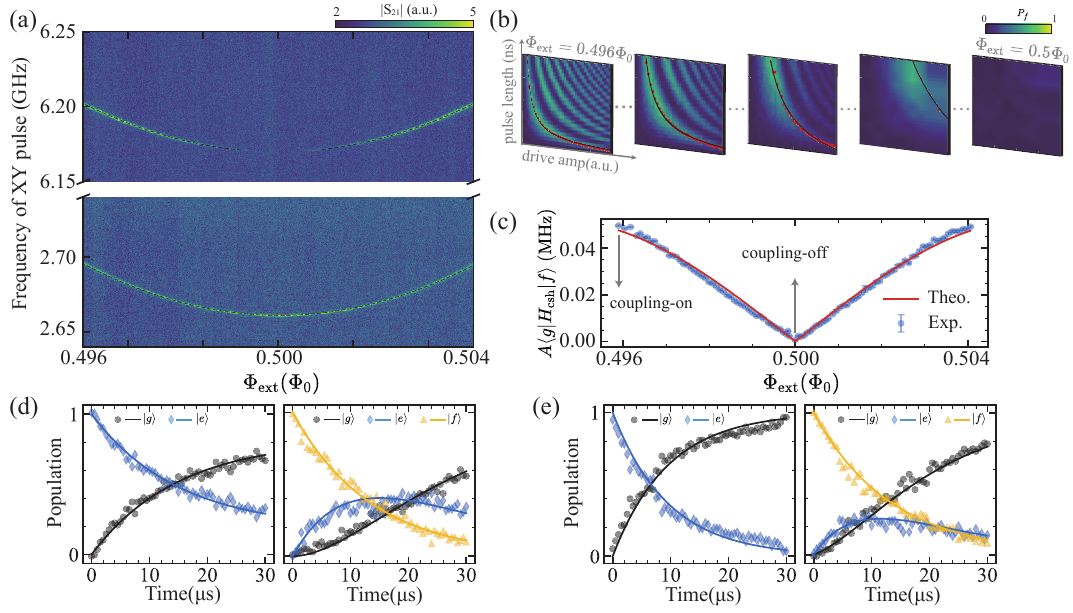}
    \caption{Properties of the C-shunt flux qubit near $0.5\Phi_0$ point. (a)~The lower plot shows the energy spectrum of the qubit's $|g\rangle \leftrightarrow |e\rangle$ transition, while the upper plot shows the energy spectrum of the $|g\rangle \leftrightarrow |f\rangle$ transition. Transition frequencies of the qubit at sweet point are $\omega_{ge}/2\pi = 2.6612~\text{GHz}, \omega_{gf}/2\pi = 6.1703~\text{GHz}$ and the qubit anharmonicity: $(\omega_{ef} - \omega_{ge})/2\pi = 0.8479~\text{GHz}$. Fitting the energy spectrum yields the qubit parameters: $C_{\text{J}} =9~\text{fF}$, $C_{\text{sh}}= 45~\text{fF}$, $\alpha =0.471$, and $I_c =88~\text{nA}$. (b)~The Rabi oscillation frequency between the $|g\rangle$ and $|f\rangle$ states at different external magnetic flux. Using these results, the transition probability as a function of the external flux can be calibrated as summarizing in (c), where $A$ is a factor related to the coupling between the XY control lines and the qubit. (d)~Population decay with time when $\Phi_{\text{ext}} = 0.5\Phi_0$. The left panel shows the qubit initially in the $|e\rangle$ state, while the right panel shows the qubit initially in the $|f\rangle$ state. From the data, we get $\Gamma_{fg} = 0.2~\text{kHz}$. (e)~Population decay with time when $\Phi_{\text{ext}} = 0.496\Phi_0$. From the data, we get $\Gamma_{fg} = 20.0~\text{kHz}$.}
    \label{fig2}
\end{figure*}
\section{Charging Model\label{sec2}}
A nondegenerate three-level QB is described by the Hamiltonian
$H_{0}=E_{g}\left|g\right\rangle\left\langle g\right| + E_{e}\left|e\right\rangle\left\langle e\right| + E_{f}\left|f\right\rangle\left\langle f\right|,$ 
where \( E_i \) is the energy of \( |i\rangle \) state with $E_{g}<E_{e}<E_{f}$. The energy of the system at time \( t \) is given by 
\(
E(t) = \mathrm{Tr}\{H_0 \rho(t)\}.
\)
Assuming the battery is initially in the ground state \( |g\rangle \), the ergotropy at time \(t\) can be expressed as 
\begin{align}
\mathcal{E}(t) = E(t) - E_g = \mathrm{Tr}\{H_0 \rho(t)\} - E_g.
\label{ergo}
\end{align}
Achieving \(\mathcal{E}_\text{max}\) requires complete population inversion from the ground state to the highest energy state. To realize such transitions, we apply a multi-frequency driving field: $H_{1}(t)=  \hbar \Omega_{ge}(t) \sin(\omega_{ge} t)\left|g\right\rangle\left\langle e\right| + \hbar \Omega_{ef}(t) \sin(\omega_{ef} t)\left|e\right\rangle\left\langle f\right| + \hbar \Omega_{gf}(t) \sin(\omega_{gf} t + \phi)\left|g\right\rangle\left\langle f\right| + \text {H.c.}$, where $\hbar$ is the reduced Planck constant and we set $\hbar = 1 $ for the remainder of the paper. $\Omega_{ge}$, $\Omega_{ef}$ and $\Omega_{gf}$ denote the amplitudes of the driving fields, $\omega_{ge}$, $\omega_{ef}$ and $\omega_{gf}$ correspond to the frequencies, respectively. Under the resonant driving condition: $\omega_{ge} = E_{e}-E_{g}, \omega_{ef} = E_{f}-E_{e}$ and $\omega_{gf} = E_{f}-E_{g}$, the interaction Hamiltonian is reduced to~\cite{dou2020}
\begin{align}\label{hamit}
H_{\mathrm{int}}(t)=  
\begin{bmatrix}
  &  0                       & \Omega_{ge}(t) & \Omega_{gf}(t)e^{i\phi}\\
  & \Omega_{ge}(t)           &   0            & \Omega_{ef}(t) \\
  &\Omega_{gf}(t)e^{-i\phi}  & \Omega_{ef}(t) & 0
\end{bmatrix}
.
\end{align}
One of the instantaneous eigenstates of the system is a dark state:
\begin{align}
    \left|E_{0}(t)\right\rangle=\frac{\Omega_{ge}(t)}{\sqrt{\Omega^2_{ge}(t) + \Omega^2_{ef}(t)}}|g\rangle-\frac{\Omega_{ef}(t)}{\sqrt{\Omega^2_{ge}(t) + \Omega^2_{ef}(t)}}|f\rangle.\nonumber
\end{align}
Under the boundary conditions $\Omega_{ge}(0)=\Omega_{ef}(\tau)=0$ and $(\Omega_{ge}(\tau),\Omega_{ef}(0))\neq(0,0)$, a slow variation of the drivings $\Omega_{ge}(t)$ and $\Omega_{ef}(t)$ keeps the system adiabatically in the dark state. As a result, the population is transferred from $|g\rangle$ to $|f\rangle$ via STIRAP~\cite{santos2019,xu2016}. The introduced CD term $\Omega_{gf}(t)e^{i\phi}\left|g\right\rangle\left\langle f\right|$ fundamentally alters this process by enabling a direct coupling between $\ket{g}$ and $\ket{f}$, thereby eliminating the need for sequential transitions through the intermediate state $\ket{e}$. This approach, known as CD-STIRAP~\cite{chen2010}, ensures that the QB remains on the adiabatic trajectory at all times, while also enhancing state coupling beyond what is achievable in the purely adiabatic case~\cite{delcampo2013,deffner2014}. Consequently, the charging power is significantly enhanced. Additionally, the inclusion of the CD pulse helps suppress detrimental non-adiabatic transitions, thereby improving the stability of the ergotropy throughout the charging process~\cite{dou2022}.

Although directly adding a CD pulse can optimize charging performance, this comparison is not entirely fair, as it results in an increase in the overall charging pulse amplitude. To properly address the optimization problem of QBs, it is essential to compare the effectiveness of different charging methods under constraint conditions, specifically by controlling the operator or trace norm of the interaction Hamiltonian~\cite{Campaioli2018}:
\begin{align}
\|H_{\text{int}}(t)\| \leq \Omega_{\text{max}},
\label{norm}
\end{align}
which is particularly evident in the entanglement-assisted charging of many-body QBs~\cite{campaioli2017,binder2015,gyhm2024}. However, this crucial aspect has often been overlooked in discussions of CD-STIRAP methods~\cite{dou2020,hu2021,ge2023,dou2022a}. Specifically, the direct inclusion of the $\Omega_{gf}(t)$ pulse would violate Eq.~(\ref{norm}), unless accompanied by a reduction in the amplitudes of the $\Omega_{ge}(t)$ and $\Omega_{ef}(t)$ pulses to maintain the overall Hamiltonian norm constraint. We will now focus on how to adjust the relative magnitudes of the CD pulse and the original STIRAP pulse under norm constraints, demonstrating that the inclusion of the CD pulse is both practical and effective.
\begin{figure*}[t]
    \centering
    \includegraphics[height = 10 cm]{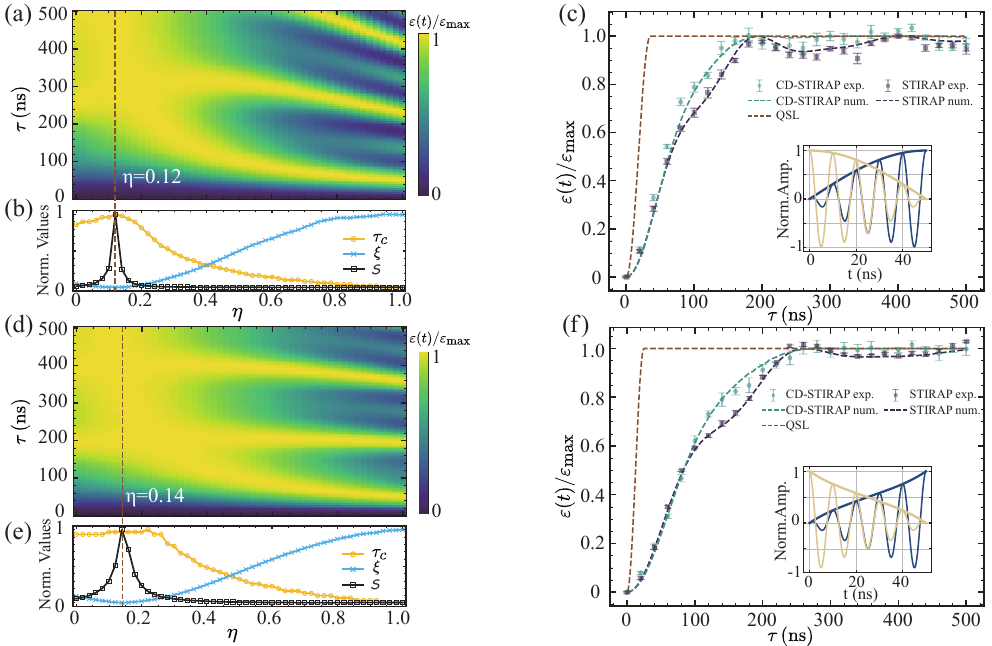}
    \caption{
    Charging optimization of CD-STIRAP method under different constraint conditions.  
    (a)~Numerical simulation of the charging process for different values of \( \eta \) under the constraint \( \Omega^2_{ge}(t) + \Omega^2_{ef}(t) + \Omega^2_{gf}(t) = \Omega^2_{\text{max}} \). The obtained values of \( \tau_c \), \( \xi \), and \( \mathcal{S} \) for different \( \eta \) values are normalized, and the results are shown in (b). The maximum value of \( S \) is obtained for \( \eta = 0.12 \). 
    (c)~Comparison of the charging curves for \( \eta = 0.12 \) and \( \eta = 0 \), with the dashed line representing the numerical simulation results. The error bars indicate standard error~(SE) of the data. All curves are within the QSL limit. The inset shows the pulse envelope of \( \Omega^{\text{tri}}_{ge}(t) \) (blue) and \( \Omega^{\text{tri}}_{ef}(t) \) (yellow). 
    (d)~Numerical simulation of the charging process for different values of \( \eta \) under the constraint \( \Omega_{ge}(t) + \Omega_{ef}(t) + \Omega_{gf}(t) = \Omega_{\text{max}} \). The maximum value of \( S \) is obtained for \( \eta = 0.14 \) as shown in (f). 
    (e)~Comparison of the charging curves for \( \eta = 0.14 \) and \( \eta = 0 \), with the dashed line representing the numerical simulation results. The inset shows the Pulse envelope of \( \Omega^{\text{cyc}}_{ge}(t) \) (blue) and \( \Omega^{\text{cyc}}_{ef}(t) \) (yellow).
    } 
    \label{fig3}
\end{figure*}
\section{C-shunt Flux Quantum Battery\label{sec3}}
Many QB platforms, such as the superconducting transmon qubit~\cite{koch2007}, exhibit extremely weak \( |g\rangle \leftrightarrow |f\rangle \) coupling due to the near-zero electric dipole interaction, posing significant challenges for implementing CD protocols. In this work, we employ the C-shunt flux qubit to address this problem. The C-shunt flux qubit consists of a loop formed by two large junctions and one small junction~\cite{yan2016}, along with a large shunt capacitor (Fig.~\ref{fig1}(c)). For the two larger junctions, both the Josephson energies and capacitances are identical, i.e., $E_{\text{J1}}=E_{\text{J2}}=E_{\text{J}}$ and $C_{\text{J1}}=C_{\text{J2}}=C_{\text{J}}$. For the small junction, the Josephson energy and capacitance are $\alpha E_{\text{J}}$ and $\alpha C_{\text{J}}$, respectively, with a scale parameter $\alpha < 0.5$ (0.471 designed in this work). The Hamiltonian of the C-shunt flux qubit is
\begin{align}
    H_{\text{csh}}=\frac{1}{2} \frac{P_{\text{p}}^{2}}{M_{\text{p}}^{2}} +\frac{1}{2} \frac{P_{\text{m}}^{2}}{M_{\text{m}}^{2}}+U\left(\varphi_{\mathrm{p}}, \varphi_{\mathrm{m}}\right),
\end{align}
with the momenta terms $P_{\sigma} = -\mathrm{i} \partial / \partial \varphi_{\sigma}~(\sigma=\mathrm{p}, \mathrm{m})$, and the mass terms $M_{\text{p}}= 2(\Phi_{\text{0}}/2\pi)^2C_{\text{J}}, M_{\text{m}}=2(\Phi_{\text{0}}/2\pi)^2C_{\text{J}}(1+2\alpha+2\beta)$, and the shunt capacitor $C_{\text{sh}} = \beta C_{\text{J}}$. The effective potential $U(\varphi_{\mathrm{p}}, \varphi_{\mathrm{m}}) = E_{\mathrm{J}}\{2(1-\cos \varphi_{\text{p}}\cos \varphi_{\text{m}})+\alpha[1-\cos (2\pi f + 2\varphi_{\text{m}})]\}$, where $\varphi_{\text{p}} = (\varphi_{1} + \varphi_{2})/2$ and $\varphi_{\text{m}} = (\varphi_{1} - \varphi_{2})/2$ are defined by the phase difference $\varphi_{1}$ and $\varphi_{2}$ across the two larger junctions~\cite{orlando1999}, and $f = \Phi_{\text{ext}}/\Phi_0$ is the external magnetic flux through the loop. When $\Phi_{\text{ext}} = 0.5\Phi_0$ (sweet spot), the potential energy of the qubit is symmetric and each quantum state is characterized by a well-defined parity~\cite{liu2005a}. Specifically, both the $|g\rangle$ and $|f\rangle$ energy levels exhibit even parity, resulting in a zero transition probability between them. However, when the external magnetic flux deviates from the sweet spot, the symmetry of the potential is broken, enabling transitions between the $|g\rangle$ and $|f\rangle$ states, i.e., the qubit is transformed from a $\Xi$-type atom to a $\Delta$-type atom~\cite{bergmann1998}. 

The device was placed in a commercial dilution refrigerator and measured at a base temperature of $10~\text{mK}$. Fig.~\ref{fig2}(a) shows the experimental two-tone energy spectrum for the \(|g\rangle \leftrightarrow |e\rangle\) and \(|g\rangle \leftrightarrow |f\rangle\) transitions of the qubit. The \(|g\rangle \leftrightarrow |f\rangle\) energy spectrum gradually disappears as the magnetic flux approaches $\Phi_{\text{ext}} = 0.5\Phi_0$, signaling that the \(|g\rangle \leftrightarrow |f\rangle\) transition probability tends to zero. 

To further investigate the relationship between the \(|g\rangle \leftrightarrow |f\rangle \) transition and $\Phi_{\text{ext}}$, we performed Rabi oscillation experiment at different bias, as shown in Fig.~\ref{fig2}(b). For this three-level system, a rectangular driving pulse with frequency \( \omega_{gf} = E_f - E_g \) and amplitude $\Omega_0$ is applied to ensure a simple linear relationship between the Rabi frequency and the amplitude, avoiding time-dependent integrals. In this case, the Hamiltonian in the rotating frame can be expressed as
\begin{align}
    H_{\text{eff}} = \Omega_0A\langle g|H_{\text{csh}}|f\rangle (|g\rangle \langle f| + \text{H.c.}),
\end{align}
where $A$ is a factor related to the coupling between the XY control lines and the qubit, and $\langle g|H_{\text{csh}}|f\rangle$ is the \(|g\rangle\leftrightarrow |f\rangle \) transition matrix element. From \( |g\rangle \) state, the time evolution of the final state $\psi(t)$ can be written as
\begin{align}
\psi(t) = & e^{-H_{\text{eff}}t}\ket{g} \\ \nonumber
= & e^{-\Omega_0A\langle g|H_{\text{csh}}|f\rangle (|g\rangle \langle f| + |f\rangle \langle g|)t}\ket{g} \\ \nonumber
= & \cos(\Omega_0A\langle g|H_{\text{csh}}|f\rangle t)|g\rangle \\ \nonumber 
  & - i\sin(\Omega_0A\langle g|H_{\text{csh}}|f\rangle t)|f\rangle.   
\end{align}
When $\psi(t)$ evolves to the \( |f\rangle \) state, we have
\begin{align}
P_f = |-i\sin(\Omega_0A\langle g|H_{\text{csh}}|f\rangle)|^2 = 1.
\end{align}
By choosing the first peak for calibration, we have $\Omega_0A\langle g|H_{\text{csh}}|f\rangle= \pi/2$. Therefore, by fitting the first peak of the Rabi oscillation experiment with an inverse proportional function, we determine the magnitude of $\langle g|H_{\text{csh}}|f\rangle$. The result in Fig.~\ref{fig2}(c) demonstrates that the \( |g\rangle \leftrightarrow |f\rangle \) transition matrix element is turned off at \(\Phi_{\text{ext}} = 0.5\Phi_0 \), and is turned on at other values, with the experimental results matching the theoretical calculations.

Additionally, we measure the qubit decay rate at different $\Phi_{\text{ext}}$, demonstrating that the dissipation rates of the qubit from energy level $|i\rangle$ to $|j\rangle$ can be controlled by $|\langle i|H_{\text{csh}}|j\rangle|^2$~\cite{blais2004a,peterer2015}, as shown in Fig.~\ref{fig2}(d) and Fig.~\ref{fig2}(e). The decay rates of the three energy levels are fitted using the following equation~\cite{peterer2015}:
\begin{align}
P_{g}(t)&=c_{1}-c_{2} e^{-\Gamma_{e g} t}+c_{3} c_{1} e^{-\left(\Gamma_{f g}+\Gamma_{f e}\right) t}, \\
P_{e}(t)&=c_{2} e^{-\Gamma_{e g} t}-c_{2} e^{-\left(\Gamma_{f g}+\Gamma_{f e}\right) t}, \\
P_{f}(t)&=c_{3} e^{-\left(\Gamma_{f g}+\Gamma_{f e}\right) t},
\end{align}
where
\begin{align}
c_{1}&=c_{2}-c_{3}\left(-\Gamma_{e g}+\Gamma_{f g}\right) /\left(\Gamma_{e g}-\Gamma_{f g}-\Gamma_{f e}\right),\\
c_{2}&=-c_{3} \Gamma_{f e} /\left(\Gamma_{e g}-\Gamma_{f g}-\Gamma_{f e}\right), \\
c_{3}&=1.
\end{align}
At $\Phi_{\text{ext}} = 0.5\Phi_0$, we found $\Gamma_{eg} = 67.0~\text{kHz}$, $\Gamma_{fe} = 71.5~\text{kHz}$, and $\Gamma_{fg} = 0.2~\text{kHz}$, where $\Gamma_{fg}$ is two order of magnitude smaller than both $\Gamma_{eg}$ and $\Gamma_{fe}$. However, at $\Phi_{\text{ext}} = 0.496\Phi_0$, the $\Gamma_{fg}$ increase due to the significant enhancement of the \( |g\rangle \leftrightarrow |f\rangle \) transition matrix element, with $\Gamma_{eg} = 104.9~\text{kHz}$, $\Gamma_{fe} = 60.0~\text{kHz}$, and $\Gamma_{fg} = 20.0~\text{kHz}$. Here, the timescale of decoherence is given by $1/\text{max}\{\Gamma_{eg},\Gamma_{fe},\Gamma_{fg}\}$. When the driving amplitude $\Omega_{\text{max}} \gg \text{max}\{\Gamma_{eg},\Gamma_{fe},\Gamma_{fg}\}$, the effects of decoherence can be neglected. In Appendix~\ref{num with dec}, we further demonstrate through numerical simulations that the above dissipation rates have a negligible impact on the subsequent charging experiments.

\section{Optimize the charging process\label{sec4}}
Now, we study an optimized charging protocol for the charging of the C-shunt flux QB under the constraint of fixed total driving amplitude. We first consider the case of square-envelope CD pulse with the sum-of-square constraint \( \Omega^2_{ge}(t) + \Omega^2_{ef}(t) + \Omega^2_{gf}(t) = \Omega^2_{\text{max}} \). Based on the principle of adiabatic quantum brachistochrone~(AQB)~\cite{rezakhani2009}, the optimal envelopes for \( \Omega_{ge}(t) \) and \( \Omega_{ef}(t) \) under adiabatic boundary conditions during $\tau$ are given by~\cite{hu2022}
\begin{align}
    \Omega^{\text{tri}}_{ge}(t) = \Omega_{\text{max}}\sin(\frac{\pi t}{2\tau}),\Omega^{\text{tri}}_{ef}(t) = \Omega_{\text{max}}\cos(\frac{\pi t}{2\tau}).
\end{align}
In our experiment, we set $\Omega_{\text{max}}/2\pi = 10\,\text{MHz}$ and define the three driving fields as 
\begin{align}
    \Omega_{ge}(t) &= \sqrt{1-\frac{\eta^2}{2}}\Omega^{\text{tri}}_{ge}(t),\\
        \Omega_{ef}(t) &= \sqrt{1-\frac{\eta^2}{2}}\Omega^{\text{tri}}_{ef}(t),\\
        \Omega_{gf}(t) &= \frac{\sqrt{2}\eta}{2}\Omega_{\text{max}},
\end{align}
where \( \eta \) is a dimensionless parameter. Fig.~\ref{fig3}(a) shows the numerical simulations of $\mathcal{E}(t)$ with different $\eta$, conducted using QuTiP~\cite{johansson2012,johansson2013}. In numerical simulations, smaller values of $\Delta \tau$ yield more accurate results under a fixed maximum charging duration. We find that $\xi$ remain robust once $\Delta \tau < 1/\Omega_{\max}$. Compared to the STIRAP method (\( \eta = 0 \)), the charging speed \( 1/\tau_c \) at \( \eta = 0.12 \) is slightly reduced, but the stability $(1/\xi)$ is significantly improved, leading to a maximum value of \( \mathcal{S} \), as illustrated in Fig.~\ref{fig3}(b). With the optimal $\eta$, the $\Omega_{ge}(t),\Omega_{ef}(t)$ and $\Omega_{gf}(t)$ are uniquely determined, enabling the experimental realization of optimal charging. In the experiment, the C-shunt flux qubit is initially biased to the sweet point ($\Phi_{\text{ext}} = 0.5~\Phi_0$) using a DC bias. A fast \( Z \)-pulse is then applied to set \(\Phi_{\text{ext}} =  0.496~\Phi_0 \), during which the driving pulses are simultaneously applied to charge the QB over a charging duration $\tau$. After charging, the fast \( Z \)-pulse and all driving pulses are turned off, returning the system to the sweet point for three-level quantum state tomography to measure the population. The ergotropy calculated using Eq.(\ref{ergo}) is presented in Fig.~\ref{fig3}(c), which shows good agreement with the numerical simulations. The inclusion of the CD pulse exhibits more stable compared to the STIRAP method. The inset of Fig.~\ref{fig3}(c) illustrates the pulse envelope of \( \Omega^{\text{tri}}_{ge}(t) \) and \( \Omega^\text{tri}_{ef}(t) \) applied during charging process. Note that numerical simulations have shown that the microwave phase $\phi = 0$ corresponds to the optimal charging process, which is also set in our experiment.

To further demonstrate the effectiveness of this method, we perform the charging optimization under the sum-of-linear constraint, with \( \Omega_{\text{max}}/2\pi = 10\,\text{MHz}\) and $\phi = 0$. The constraint condition is written as \( \Omega_{ge}(t) + \Omega_{ef}(t) + \Omega_{gf}(t) = \Omega_{\text{max}} \). In this case, the optimal envelope for \( \Omega_{ge}(t) \) and \( \Omega_{ef}(t) \) is the arc of the cycloid~\cite{hu2022}
\begin{align}
        \Omega^{\text{cyc}}_{ge}(t) & = \frac{\Omega_{\max }}{2}\left[1-\tan \left(\frac{\pi(1-2 t / \tau)}{4}\right)\right],\\ \nonumber
        \Omega^{\text{cyc}}_{ef}(t) &  = \Omega_{\max }-\Omega^{\text{cyc}}_{ge}(t).
\end{align}
The three driving fields are defined as
\begin{align}
        \Omega_{ge}(t)  &= (1 - \frac{\eta}{2}) \Omega^{\text{cyc}}_{ge}(t),\\
        \Omega_{ef}(t)  &= (1 - \frac{\eta}{2}) \Omega^{\text{cyc}}_{ef}(t),\\
        \Omega_{gf}(t) &= \frac{\eta}{2}\Omega_{\text{max}}.
\end{align}
Figures \ref{fig3}(d) and \ref{fig3}(e) display the numerical simulations for different $\eta$, where the optimal charging process corresponds to \( \eta = 0.14 \). Figure.~\ref{fig3}(f) presents the experimental results, the ergotropy $\mathcal{E}(t)$ with the added CD pulse demonstrates more stable behavior after the system is fully charged.

It is worth noting that, in our experiment, neither the sum-of-square nor the sum-of-linear results exhibit a significant acceleration of the stable QB charging process by CD-STIRAP compared to STIRAP. Instead, an enhancement of stability at small amplitudes is observed. This suggests, on the one hand, that the acceleration effect is strongly correlated with the increase in the total drive amplitude, which was suppressed in our experiment. On the other hand, it may also stem from the fact that the square-shaped CD pulse employed here is not its optimal form. Further work could focus on deriving the optimal pulse envelope under the constraint of limited total amplitude.
\begin{figure}[t]
    \centering
    \includegraphics[height = 5 cm]{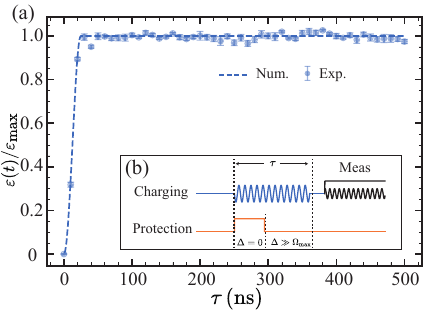}
    \caption{A charging process achieving QSL. (a)~Blue points represent experimental data, and the dashed line represents the numerical simulation results. (b)~Schematic of the pulse sequence used in this method. The detuning between the drive frequency and the $g$–$f$ transition frequency is denoted by $\Delta$, where $\Delta = (E_f - E_g) - \omega_{gf}$. The protective pulse leads to a resonant cutoff when $\Delta = 0$, and a detuned cutoff when $\Delta / 2\pi = 48~\mathrm{MHz}$.} 
    \label{fig4}
\end{figure}
\section{Charging scheme reaching the quantum speed limit\label{sec5}}
Finally, based on the energy level structure of the C-shunt flux QB, we implement a charging scheme that reaches the quantum speed limit (QSL)~\cite{deffner2013}. The QSL sets a fundamental lower bound on the time required for a system to transition between states and is generally expressed by ($\hbar = 1$)
\begin{align}
T(|\psi\rangle,|\phi\rangle)= \frac{\arccos |\langle\psi \mid \phi\rangle|}{\min \{E, \Delta E\}},
\end{align}
where $|\psi\rangle$ is the initial state and $|\phi\rangle$ is the final state, while $E$ and $\Delta E$ correspond to the time-averaged energy and standard deviation of the Hamiltonian $H_{\text{int}}(t)$. According to the derivation in Appendix~\ref{qsl}, the state transfer from $\ket{g}$ to $\ket{f}$ can reach the QSL $T(\ket{g}, \ket{f}) = \pi/(2\Omega_{\max})$ only when $\Omega_{ge}(t)=0$, $\Omega_{ef}(t)=0$, and $\Omega_{gf}(t)=\Omega_{\max}$. It requires $\Omega_{gf}(t)$ to remain at its maximum value throughout the entire evolution.

The specific implementation involves using a fast Z-pulse to assist the charging process, ensuring that $\omega_{gf} = E_f - E_g$ only for a limited duration. In our scheme, the $Z$-pulse adopts a square-envelope shape to guarantee that $\Omega_{gf}(t)$ remains at its maximum value throughout the evolution, with a duration fixed at $\tau = \pi/(2\Omega_{\max})$. This effectively shifts the prior specification of the charging-pulse duration to the $Z$-channel, where it acts as a protection pulse, so that no predefined charging-pulse duration is required on the charging channel. This strategy guarantees that the qubit ceases to be driven once it is fully charged, thereby preserving the maximum ergotropy. The experimental results shown in Fig.~\ref{fig4}(a) obtained with \( \Omega_0/2\pi = 10~\text{MHz} \) are in good agreement with the theoretical predictions. The charging time is measured to be \( \tau_c = T(|g\rangle,|f\rangle) = 25~\text{ns}\). This method underscores the advantages of the C-shunt flux qubit as a three-level QB, highlighting its promising potential for future applications.
\section{Conclusion\label{sec6}}
In conclusion, we achieve optimal charging on the C-shunt flux qubit by utilizing its non-forbidden $|g\rangle \leftrightarrow |f\rangle$ transition. To contain the Hamiltonian norm, we impose constraints that allow us to harness the advantages of STIRAP without increasing the driving strength. By introducing the $\mathcal{S}$ parameter, we effectively quantify and unify the speed and stability of the charging curve. Furthermore, we demonstrate stable charging that reaches the QSL in the three-level quantum battery through flexible switching of the $|g\rangle \leftrightarrow |f\rangle$ coupling. These methods hold significant implications for the future optimization of quantum battery charging. We also note that while square-wave CD pulses are employed to better meet the constraint conditions, the form of the CD pulse is analytically determined in some CD-STIRAP studies~\cite{chen2010,dou2022}. Future work could focus on unifying the envelopes of the three driving fields via AQB theory~\cite{rezakhani2009} to explore potentially more optimal charging trajectories. This would further emphasize the unique role of quantum batteries in both quantum thermodynamics~\cite{vinjanampathy2016quantum,kosloff2013quantum} and quantum optimal control~\cite{mahesh2023quantum,werschnik2007quantum}, while advancing their practical implementation~\cite{albarelli2020,chiribella2017,campbell2025}.
\section{ACKNOWLEDGMENTS}
This work was supported by the Micro/Nano Fabrication Laboratory of Synergetic Extreme Condition User Facility (SECUF). The devices were made at the Nanofabrication Facilities at the Institute of Physics, CAS in Beijing. This work was supported by Innovation Program for Quantum Science and Technology
(Grant No. 2021ZD0301800), the National Natural Science Foundation of China (Grants No. 12204528, No. 92265207, No. T2121001, No. 92065112, No. 92365301, No. T2322030, No. 12574540, No. 12504593). We acknowledge the state key laboratory of high performance computing.
\appendix
\section{Device fabrication}

The device fabrication was performed using micro-nano technology. Initially, a 100-$\text{nm}$-thick aluminum layer was deposited through electron beam evaporation (Adnano JEB-4). Before deposition, the substrate was baked at 200 degrees for 3 hours to remove water vapor. Subsequently, the primary circuit components, including resonant cavities, transmission lines, and capacitors, were defined using laser direct writing (DWL66+) followed by wet chemical etching to achieve the desired geometries. The fabrication of Josephson junctions was performed using electron beam lithography (EBPG5200) combined with double-angle evaporation (Adnano JEB-4), which represents a crucial step for precisely controlling the parameter $\alpha$. A number of airbridges have been fabricated around the transmission line to balance the electric potential on both sides and suppress slotline modes. Finally, the fully fabricated chip was packaged in a copper sample holder via aluminum wire bonding, ensuring robust electrical connections suitable for low-temperature measurements.

The information of the device is listed in Tab.~\ref{parameters}. Additionally, the large positive anharmonicity ($\omega_{ef} - \omega_{ge}$) of the C-shunt flux qubit offers two key advantages: the higher energy of the $\ket{f}$ state enhances the maximum ergotropy $\mathcal{E}_\text{max}$, while simultaneously suppressing leakage to undesired levels during charging, enabling more precise operations.

\begin{figure*}[t]
    \centering
    \includegraphics[scale = 0.9]{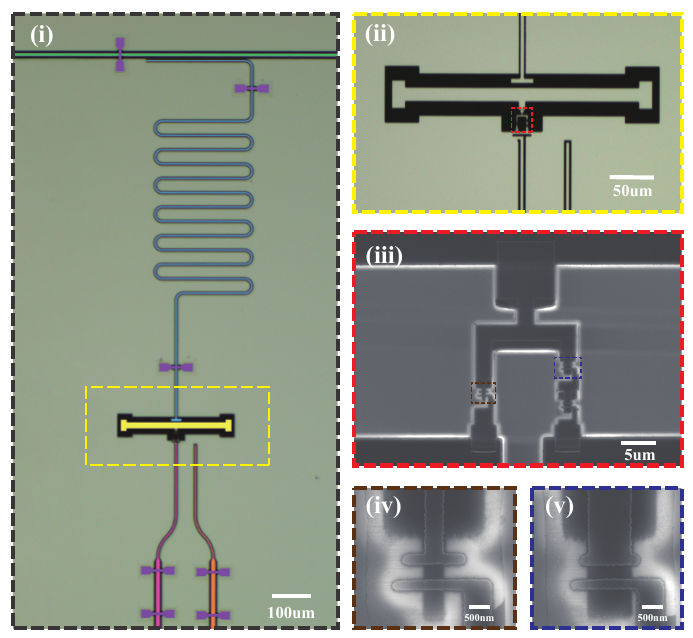}
    \caption{Photograph of the device. (i)~False-color micrograph of the device. The circuit elements are color-coded as follows: readout line (green), readout resonator (blue), shunt capacitor (yellow), XY control line (orange), Z control line (pink) and airbridges (purple). (ii)~Localized micrograph of the shunt capacitor, with the qubit loop outlined in red. (iii)–(iv)~Scanning electron micrographs: (iii)~qubit loop, (iv)~small Josephson junction, (v)~Large Josephson junction.} 
    \label{device}
\end{figure*}

Fig.~\ref{device} presents a photograph of the device. The device is shown in a false-color micrograph in (i), where different circuit elements are highlighted for clarity: the readout line (green), readout resonator (blue), shunt capacitor (yellow), XY control line (orange), Z control line (pink), and airbridges (purple). The region enclosed by the yellow box in Fig.~\ref{device}(i) is magnified in (ii), providing a closer view of the shunt capacitor, with the qubit loop outlined in red for emphasis. This qubit loop, in turn, is further enlarged in (iii), where the brown box highlights a small Josephson junction, shown in greater detail in (iv), while the blue box marks a large Josephson junction, depicted in (v). 
\section{The Hamiltonian of capacitively shunted flux qubit}

\begin{figure*}[t]
    \centering
    \includegraphics[scale = 1.9]{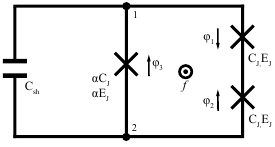}
    \caption{Schematic circuit diagram of capacitively shunted flux qubit. The nodes 1 and 2 represent the superconducting islands, while $f$ represents total external flux in the loop. Josephson junctions 1 and 2 both have Josephson energies $E_{\text{J}}$ and capacitance $C_{\text{J}}$ and Josephson junction 3 has $\alpha$ times the capacitance and Josephson energy of 1 and 2. $C_{\text{sh}}$ is the shunt capacitor. } 
    \label{qubit}
\end{figure*}

The circuit of the capacitively shunted flux qubit is shown in Fig.~\ref{qubit}. The total external flux in the loop is $f$ in units of the flux quantum $\Phi_\text{0}$. According to the condition of magnetic fluxion, we have
\begin{align}
    \varphi _{1} -\varphi _{2} +\varphi _{3} =2\pi f, 
\end{align}
where $\varphi _{1}, \varphi _{2}, \varphi _{3}$ is the magnetic fluxion of Josephson junction 1, 2 and 3, respectively. The total Josephson energy $U$ can be written as
\begin{align}
U({\varphi})= & \sum_iE_{\text{J}i}(1-\cos\varphi_\text{i})\\\nonumber
= & E_\mathrm{{J}}(1-\cos\varphi_{1}) + E_\mathrm{{J}}(1-\cos\varphi_{2})\\\nonumber
& +\alpha E_\mathrm{{J}}( 1-\cos\varphi_{3})\\\nonumber
=& E_\mathrm{{J}}(2+\alpha-\cos \varphi_{1}-\cos \varphi_{2} \\ \nonumber
 & -\alpha \cos \left(2 \pi f+\varphi_{1}-\varphi_{2}\right)), 
\end{align}
and the kinetic energy part of the Hamiltonian is
\begin{align}
\vec{T}=\frac{1}{2}\vec{P}^{T}\cdot\mathbf{M}^{-1} \cdot\vec{P},
\end{align}
where $\mathbf{M}=(\Phi_\text{0}/2\pi)^2\mathbf{C}$ is the effective mass. The momenta $\vec{P}$ and the capacitance matrix $\mathbf{C}$ are given by\cite{orlando1999}
\begin{align}
\vec{P} &=-i\hbar\begin{pmatrix}
 \frac{\partial}{\partial \varphi_{1}}  \\ \frac{\partial}{\partial \varphi_{2}}
\end{pmatrix},\\
\mathbf{C} &= C_\mathrm{J}\begin{pmatrix}
  1+\alpha +\beta &-(\alpha +\beta) \\
  -(\alpha +\beta)&1+\alpha +\beta
\end{pmatrix}.
\end{align}
The reduced Hamiltonian $H_{csh}$ is obtained by choosing $\varphi_\text{p} =(\varphi_\text{1}+\varphi_\text{2})/2$ and $   
\varphi_\text{m} =(\varphi_\text{1}-\varphi_\text{2})/2$ as coordinates
\begin{align}
    H_{\text{csh}}=\frac{1}{2} \frac{P_{\text{p}}^{2}}{M_{\text{p}}^{2}} +\frac{1}{2} \frac{P_{\text{m}}^{2}}{M_{\text{m}}^{2}}+U\left(\varphi_{\mathrm{p}}, \varphi_{\mathrm{m}}\right),
\end{align}
where $P_{\text{p}} = -\mathrm{i} \partial / \partial \varphi_{\text{p}}$, 
    $P_{\text{m}} = -\mathrm{i} \partial / \partial \varphi_{\text{m}}$, $M_{\text{p}} = 2(\Phi_{\text{0}}/2\pi)^2C_{\text{J}}$, $M_{\text{m}} =2(\Phi_{\text{0}}/2\pi)^2C_{\text{J}}(1+2\alpha+2\beta)$ and the shunt capacitor is $C_{\text{sh}} = \beta C_{\text{J}}$. The effective potential is 
\begin{align}
U(\varphi_{\mathrm{p}}, \varphi_{\mathrm{m}}) = &  E_{\mathrm{J}}\{2(1-\cos \varphi_{\text{p}}\cos \varphi_{\text{m}}) \\ \nonumber
& +\alpha[1-\cos (2\pi f + 2\varphi_{\text{m}})]\}.  
\end{align}
Following the approach in Ref.~\cite{yan2016}, we now derive the effective Hamiltonian under the three-level approximation. The presence of the large shunt capacitance ensures leaving an effective one-dimensional Hamiltonian
\begin{align}ing
H_{m} = \frac{1}{2}\frac{P_{m}^{2}}{M_{m}} + U_{m}(\varphi_{m}),
\end{align}
where
\begin{align}
U_{m}(\varphi_{m}) = E_{J}\{-2\cos\varphi_{m} + \alpha\cos(2\pi f_{b} + 2\varphi_{m})\},
\end{align}
and $f_{b}=f-1/2$ is the reduced external flux bias.

The potential $U_{m}(\varphi_{m})$ has a minimum at $\varphi_{m}^{\ast}$, defined by
\begin{align}
\frac{\partial U_{m}}{\partial \varphi_{m}}\Big|_{\varphi_{m}=\varphi_{m}^{\ast}}=0.
\end{align}
Expanding around this minimum with $\varphi_{m}'=\varphi_{m}-\varphi_{m}^{\ast}$ gives
\begin{align}
U_{m}(\varphi_{m}) \approx U^{(0)}
+ \frac{1}{2}U^{(2)}(\varphi_{m}')^{2}
+ \cdots ,
\end{align}
where $U^{(k)}$ are the $k$th derivatives of $U_{m}$ evaluated at $\varphi_{m}^{\ast}$.

At the quadratic level, the Hamiltonian becomes
\begin{align}
H_{0} = \frac{1}{2}\frac{P_{m}^{2}}{M_{m}} + \frac{1}{2}U^{(2)}(\varphi_{m}')^{2}.
\end{align}
This is quantized by introducing bosonic operators
\begin{align}
\hat{\varphi}_{m}' = \varphi_{Z}(\hat b+\hat b^{\dagger}), \qquad
\hat P_{m} = \frac{i}{2\varphi_{Z}}(\hat b^{\dagger}-\hat b),
\end{align}
where the zero-point fluctuation is defined as
\begin{align}
\varphi_{Z} = \Big(\frac{1}{2M_{m}U^{(2)}}\Big)^{1/4}.
\end{align}
In this representation the quadratic Hamiltonian is
\begin{align}
H_{0} = \hbar \Omega_{m}^{(0)}\Big(\hat b^{\dagger}\hat b + \tfrac{1}{2}\Big),
\end{align}
with the harmonic frequency
\begin{align}
\hbar \Omega_{m}^{(0)} = \sqrt{\frac{U^{(2)}}{M_{m}}}.
\end{align}

The higher-order terms provide anharmonic corrections. To leading order, the quartic term gives a Kerr-type contribution
\begin{align}
V \simeq \frac{U^{(4)}\varphi_{Z}^{4}}{8}\, \hat b^{\dagger}\hat b^{\dagger}\hat b\hat b.
\end{align}

Restricting the system to the lowest three eigenstates $\{|g\rangle, |e\rangle, |f\rangle\}$, the effective Hamiltonian can be written as
\begin{align}
H_{\text{eff}} = E_{g}|g\rangle\langle g|
+ E_{e}|e\rangle\langle e|
+ E_{f}|f\rangle\langle f|,
\end{align}
with approximate level energies
\begin{align}
E_{g} \approx 0,\qquad
E_{e} \approx \hbar\Omega_{m}^{(0)},\qquad
E_{f} \approx 2\hbar\Omega_{m}^{(0)} + \frac{U^{(4)}\varphi_{Z}^{4}}{4}.
\end{align}


\begin{table*}[t]
    \centering
    \caption{The parameters of the qubit.}
    \renewcommand{\arraystretch}{1.2} 
    \begin{tabular}{cccccc}
        \hline
        $\Phi_{\text{ext}}$& Parameter & Symbol &  Unit & Value \\
        \hline
        \multirow{12}{*}{$0.5\Phi_0$}
        &Frequency of readout resonator & $\omega_r/2\pi$ &  $\text{GHz}$ & 6.4214 \\
        &Coupling strength between qubit and readout resonater & $g_{qr}/2\pi$ &  $\text{MHz}$ & 29.8 \\
        &Shunt capacitor & $C_\text{sh}$  &  $\text{fF}$ & 45 \\
        & Ratio of large capacitor to small capacitor & $\alpha$ &  & 0.471 \\   
        &Transition frequency of $|g\rangle \leftrightarrow |e\rangle$ &$\omega^{ge}_q/2\pi$   &  $\text{GHz}$ & 2.6612 \\
        &Transition frequency of $|g\rangle \leftrightarrow |f\rangle$& $\omega^{gf}_q/2\pi$   &  $\text{GHz}$ & 6.1703 \\
        &Qubit anharmonicity & $\eta/2\pi$ &  $\text{GHz}$ & 0.8479 \\
        & Decay rate from $|e\rangle$ to $|g\rangle$ & $\Gamma_{eg}$   &     $\text{kHz}$   &  67.0 \\
        & Decay rate from $|f\rangle$ to $|e\rangle$ & $\Gamma_{fe}$   &     $\text{kHz}$   &  71.5 \\
        & Decay rate from $|f\rangle$ to $|g\rangle$ & $\Gamma_{fg}$   &     $\text{kHz}$   &  0.2 \\
        & Ramsey dephasing time  & $T_2^*$ &  $\mu s$ & 6.3\\
        & Spin-echo dephasing time & $T_2^{echo}$ & $\mu s$ & 23.4 \\
        \hline
        \multirow{6}{*}{$ 0.496\Phi_0$}
        &Transition frequency of $|g\rangle \leftrightarrow |e\rangle$ & $\omega^{ge}_q/2\pi$   &  $\text{GHz}$ & 2.7123 \\
        &Transition frequency of $|g\rangle \leftrightarrow |f\rangle$ & $\omega^{gf}_q/2\pi$   &  $\text{GHz}$ & 6.2180 \\
        & Decay rate from $|e\rangle$ to $|g\rangle$ & $\Gamma_{eg}$   &     $\text{kHz}$   &  104.9 \\
        & Decay rate from $|f\rangle$ to $|e\rangle$ & $\Gamma_{fe}$   &     $\text{kHz}$   &  60.0 \\
        & Decay rate from $|f\rangle$ to $|g\rangle$ & $\Gamma_{fg}$   &     $\text{kHz}$   &  20.0 \\
        &Qubit anharmonicity & $\eta/2\pi$ &  $\text{GHz}$ & 0.7934 \\
        \hline
    \end{tabular}
    \label{parameters}
\end{table*}

\section{Measurement setup}
The low-temperature attenuation and filtering of the control lines, as well as the readout configuration in the measurement system, are consistent with those presented in our previous work~\cite{ruan2024}. In this study, we apply three simultaneous microwave pulses: $\Omega_{ge}(t)\sin(\omega_{ge}t), \ \Omega_{fe}(t)\sin(\omega_{ef}t), \ \text{and} \ \Omega_{ge}(t)\sin(\omega_{gf}t),$ where $\omega_{gf}$ is approximately twice $\omega_{ef}$ and $\omega_{ge}$. In a conventional IQ mixing scheme, this setup exceeds the output range of the arbitrary waveform generator~(AWG). As a result, two sets of microwave sources~(MWs), AWGs, and IQ mixers are required for frequency mixing. The first set generates $\omega_{ge}$ and $\omega_{ef}$ by setting the local oscillator (LO) frequency of the microwave source between $\omega_{ge}$ and $\omega_{ef}$, enabling the AWG to produce pulses at approximately 400 MHz. The second set generates pulses at the frequency $\omega_{gf}$. All signals are then combined at room temperature. 

We explored two methods for signal combining and control. The first method, shown in Fig.~\ref{measure setup}a, uses a microwave combiner to combine the pulses generated by two IQ mixers. These combined pulses are then passed through low-temperature attenuation and filtering before being routed to the qubit's XY control line. The fast-Z pulses output from the AWG are combined with DC signals at low temperature using a bias tee, with the magnetic flux bias applied through the qubit's Z control line.

The second method, shown in Fig.~\ref{measure setup}b, involves merging the combined XY control signal with the fast-Z signal at room temperature using a directional coupler. The signals are then combined with DC at low temperature through a bias tee and input through the qubit's Z control line.

Due to the capacitive and inductive coupling between the Z control line and the qubit, this XY-Z combining approach can effectively control the qubit while also reducing the cost of low-temperature wiring. In our experiments, both methods yielded equivalent results, demonstrating that the XY-Z combining scheme does not introduce additional experimental errors. However, a major source of experimental error stemmed from pulse distortion introduced by the microwave combiner, which led to envelope distortion of the CD and STIRAP pulses, causing discrepancies between the experimental results and numerical simulations.

\section{Numerical Simulations With Decoherence}\label{num with dec}
\begin{figure*}[t]
    \centering
    \includegraphics[scale = 0.8]{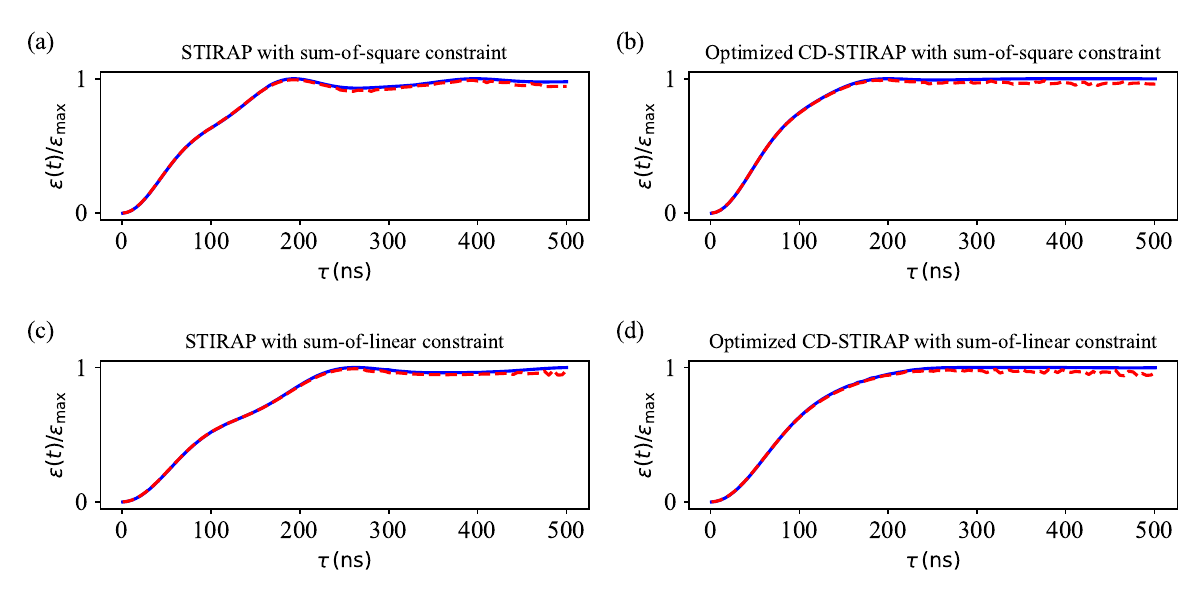}
    \caption{The effect of decoherence on the charging process of the QB. Numerical simulations including decoherence with parameters $\Gamma_{eg} = 104.9~\text{kHz}$, $\Gamma_{fe} = 60.0~\text{kHz}$, and $\Gamma_{fg} = 20.0~\text{kHz}$ are shown as red dashed lines, while the ideal case without decoherence is represented by blue solid lines.(a)~Numerical simulations of the STIRAP charging process with sum-of-square constraint. (b)~Numerical simulations of the CD-STIRAP charging process with sum-of-square constraint, $\eta = 0.12$. (c)~Numerical simulations of the STIRAP charging process with sum-of-linear constraint. (d)~Numerical simulations of the CD-STIRAP charging process with sum-of-linear constraint, $\eta = 0.14$.} 
    \label{dec}
\end{figure*}

\begin{figure*}[t]
    \centering
    \includegraphics[scale = 0.55]{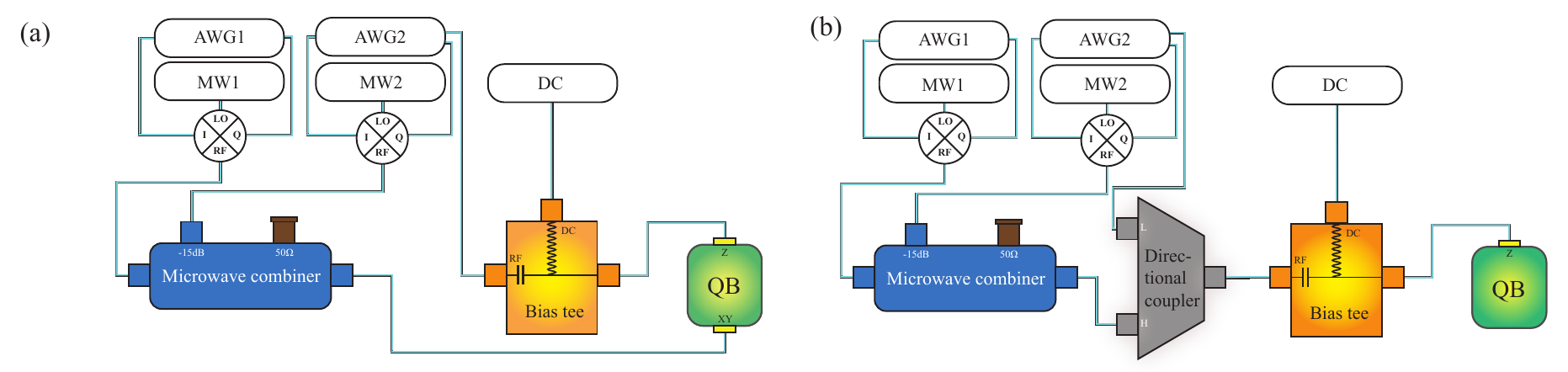}
    \caption{The two methods of driving signal combining and control.} 
    \label{measure setup}
\end{figure*}

In Fig.~\ref{dec}, we present numerical simulations of STIRAP and CD-STIRAP under two different constraints, taking into account decoherence effects. The decoherence parameters are chosen according to the experimental values at $\Phi_{\text{ext}} = 0.496\Phi_0$: $\Gamma_{eg} = 104.9~\text{kHz}$, $\Gamma_{fe} = 60.0~\text{kHz}$, and $\Gamma_{fg} = 20.0~\text{kHz}$. The red dashed curves represent the simulations including decoherence, while the blue solid curves correspond to the ideal case without decoherence. As can be seen, the two results are in close agreement. This demonstrates that under the lifetime conditions of our quantum battery, decoherence effects can be safely neglected, thereby justifying the use of unitary operations to describe the charging process.
\section{Three level quantum state tomography}
In this study, ergotropy is obtained from the density matrix $\rho(t)$ through three-level quantum state tomography (QST) measurements of the qutrit~\cite{thew2002}. The three-level tomography measurement basis $\psi_i$, rotation operator $U_i$ and corresponding unitary matrix $\lambda_i$ used in the experiment are listed in Tab.~\ref{tomo}. After obtaining the expectation values of the measured operators, the density matrix is reconstructed using maximum likelihood estimation~\cite{james2001}.

\begin{table*}[t]
    \centering
    \caption{Complete set of rotation operations and corresponding measurement basis for three-level qutrit state tomography, where $j$ denotes the imaginary unity.}
    \renewcommand{\arraystretch}{3.5} 
    \renewcommand{\tabcolsep}{15pt} 
    \begin{tabular}{cccc}
        \hline
        $i$ & $U_i$ & $\psi_i$ & $\lambda_i$  \\
        \hline
1 & I & $|g\rangle$ & \makecell{$\begin{bmatrix}1 & 0 & 0 \\ 0 & 1 & 0 \\ 0 & 0 & 1\end{bmatrix}$} \\
2 & $(\pi/2)_x^{ge}$ & $(|g\rangle - j|e\rangle)/\sqrt{2}$ & \makecell{$\frac{1}{\sqrt{2}}\begin{bmatrix}1 & -j & 0 \\ -j & 1 & 0 \\ 0 & 0 & \sqrt{2}\end{bmatrix}$} \\
3 & $(\pi/2)_y^{ge}$ & $(|g\rangle + |e\rangle)/\sqrt{2}$ & \makecell{$\frac{1}{\sqrt{2}}\begin{bmatrix}1 & -1 & 0 \\ 1 & 1 & 0 \\ 0 & 0 & \sqrt{2}\end{bmatrix}$} \\
4 & $(\pi)_x^{ge}$ & $|e\rangle$ & \makecell{$\begin{bmatrix}0 & -j & 0 \\ -j & 0 & 0 \\ 0 & 0 & 1\end{bmatrix}$} \\
5 & $(\pi/2)_x^{ef}$ & $|g\rangle$ & \makecell{$\frac{1}{\sqrt{2}}\begin{bmatrix}\sqrt{2} & 0 & 0 \\ 0 & 1 & -j \\ 0 & -j & 1\end{bmatrix}$} \\
6 & $(\pi/2)_y^{ef}$ & $|g\rangle$ & \makecell{$\frac{1}{\sqrt{2}}\begin{bmatrix}\sqrt{2} & 0 & 0 \\ 0 & 1 & -1 \\ 0 & 1 & 1\end{bmatrix}$} \\
7 & $(\pi)_x^{ge}(\pi/2)_x^{ef}$ & $(|e\rangle - j|f\rangle)/\sqrt{2}$ & \makecell{$\frac{1}{\sqrt{2}}\begin{bmatrix}0 & \sqrt{2} & 0 \\ 1 & 0 & -j \\ -j & 0 & 1\end{bmatrix}$} \\
8 & $(\pi)_x^{ge}(\pi/2)_y^{ef}$ & $(|e\rangle + |f\rangle)/\sqrt{2}$ & \makecell{$\frac{1}{\sqrt{2}}\begin{bmatrix}0 & \sqrt{2} & 0 \\ 1 & 0 & -1 \\ 1 & 0 & 1\end{bmatrix}$} \\
9 & $(\pi)_x^{ge}(\pi)_x^{ef}$ & $|f\rangle$ & \makecell{$\begin{bmatrix}0 & 0 & j \\ -j & 0 & 0 \\ 0 & -j & 0\end{bmatrix}$} \\

        \hline
    \end{tabular}
    \label{tomo}
\end{table*}

\section{Phase of CD pulse}
In our numerical simulations, we demonstrate that the optimal charging process corresponds to a phase of \( \Omega_{gf}(t) \) equal to zero, as shown in Fig.~\ref{phase}. In both the sum-of-square and sum-of-linear constraints, the ergotropy of the charging evolution varies periodically with \( \phi \) as \( 2\pi \) when \( \eta \) is set to its optimal value. The maximum \( \mathcal{S} \) corresponds to a phase of \( \phi = 0 \).

\section{Validation of Adiabaticity}
\begin{figure*}[h]
        \centering
        \includegraphics[width=0.8\linewidth]{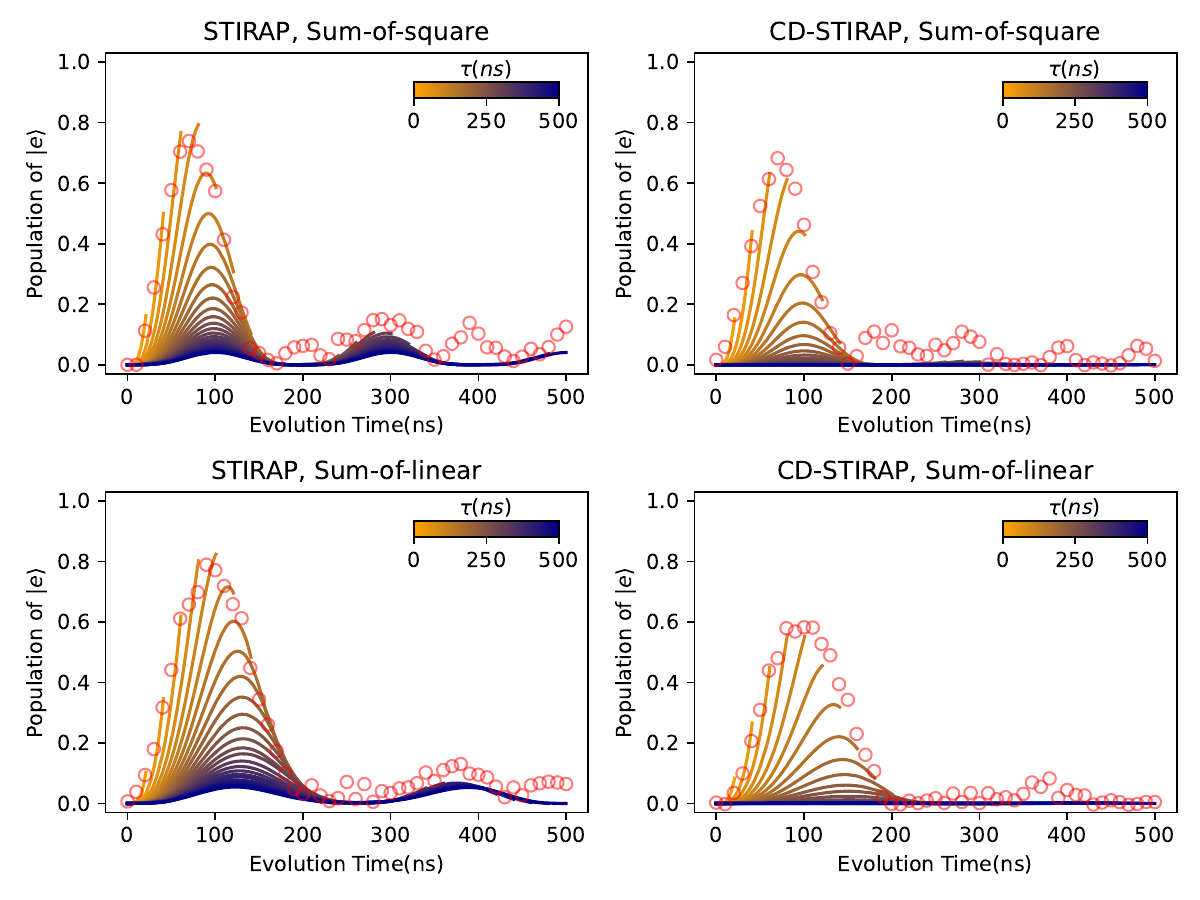}
        \caption{The population of the intermediate state $|e\rangle$ as a function of both the charging time $\tau$ and the evolution time. As $\tau$ increases from short to long, the color of the curves changes from orange to blue. The red open circles represent the experimental data.}
        \label{p1data}
    \end{figure*}

\begin{figure*}[t]
    \centering
    \includegraphics[scale = 0.7]{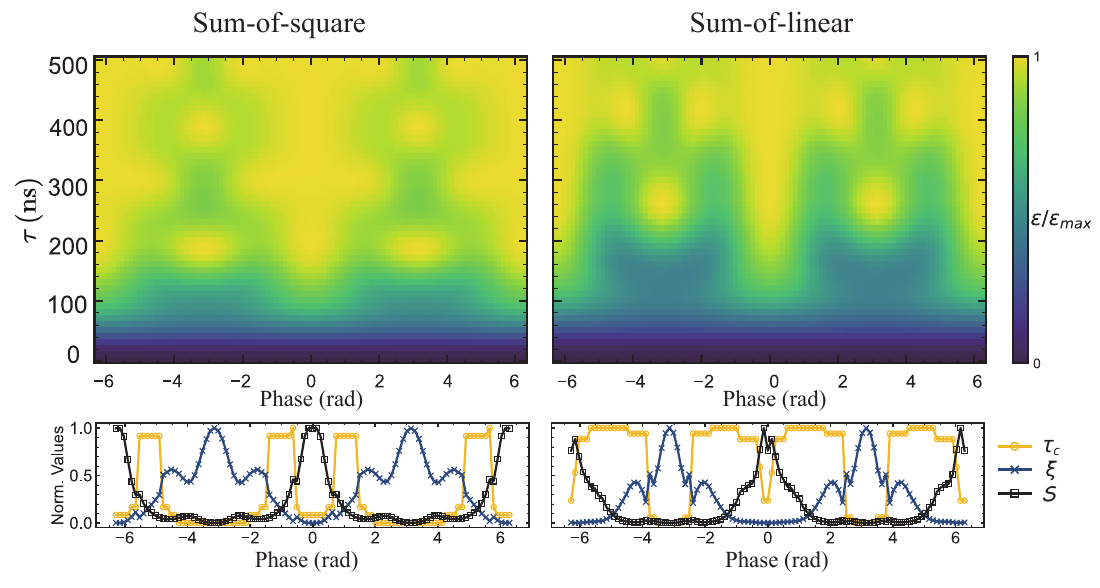}
    \caption{The numerical simulations of ergotropy during charging in CD-STIRAP with different  \( \phi  \) in \( \Omega_{gf}(t) \).} 
    \label{phase}
\end{figure*}

To examine whether adiabaticity is satisfied in our experiment, we use the population of the intermediate state $\lvert e\rangle$ as an indicator. In the measurements, $\lvert e\rangle$ was not recorded at each point during the evolution, but only at the final time $\tau$. To address this, we performed numerical simulations. Figure~\ref{p1data} shows the $\lvert e\rangle$ population during STIRAP and CD-STIRAP as a function of charging time $\tau$. The color of the traces changes from orange to blue as $\tau$ increases. Two features are evident. Firstly, for short $\tau$, the adiabatic condition is not fully satisfied and $\lvert e\rangle$ becomes significantly populated, but long $\tau$ suppresses its occupation. Secondly, $\lvert e\rangle$ remains nearly unpopulated after the addition of the CD pulse. 
The experimental $|e\rangle$ data are plotted as the red open circles. In principle, the experiment data follow the envelope of the numerical simulation curves. These results confirm that adiabaticity is not fully satisfied at short $\tau$, but is achieved at long $\tau$. It also demonstrates that the CD pulse effectively reduces the population of the intermediate state.

\section{Thermodynamic cost}
Now, we consider the change in thermodynamic cost after adding the CD pulse compared to the STIRAP method. During the charging process, the energy $\Sigma_{\text{total}}$ from the external fields can be divided into an absorbed component $\Sigma_{\text{abs}}$ and a cost component $\Sigma_{\tau}$. According to the definition in Ref.~\cite{deffner2021}, the thermodynamic cost in a unitary process is defined as
\begin{align}
    \Sigma_{\tau}=\frac{1}{\tau} \int_{0}^{\tau}\|H_{\text{int}}(t)\| \mathrm{d} t,
\end{align}
where \( \|H_{\text{int}}(t)\| = \sqrt{\text{tr}(H_{\text{int}}^2(t))} \) is the instantaneous Hilbert–Schmidt norm of $H_{\text{int}}$. In our protocol, the thermodynamics cost reads
\begin{align}
    \Sigma_{\tau}=\frac{\sqrt{2}\hbar}{\tau} \int_{0}^{\tau}\sqrt{\Omega^2_{ge}(t) + \Omega^2_{ef}(t) + \Omega^2_{gf}(t)} \mathrm{d} t.
    \label{cost}
\end{align}
Using the above expression, the thermodynamic efficiency of the charging process can be defined as
\begin{align}
    \mu = \frac{\mathcal{E}_{\text{max}}}{\Sigma_{\text{total}}} = \frac{\omega_{ge} + \omega_{ef}}{\omega_{ge} + \omega_{ef} + \Sigma_{\tau}/\hbar}.
\end{align}

As shown in Eq.~\ref{cost}, under the sum-of-square constraint, the addition of the CD pulse does not alter the thermodynamic efficiency which remains $\mu^{\text{STIRAP}}_{\text{tri}} = \mu^{\text{CD-STIRAP}}_{\text{tri}} \approx 99.772\% $. On the other hand, for the case of sum-of-linear constraint, the efficiency of STIRAP protocol is given by $\mu^{\text{STIRAP}}_{\text{cyc}}\approx 99.815\% $, and the efficiency of CD-STIRAP protocol is given by $\mu^{\text{CD-STIRAP}}_{\text{cyc}}\approx 99.809\% $.
From these results, it can be seen that while the thermodynamic efficiencies of the methods are numerically close, their charging speeds and stability differ considerably. Therefore, thermodynamic efficiency alone does not fully characterize the charging performance of the QB, highlighting the necessity of the proposed $\mathcal{S}$ metric.

\section{Derivation of the Quantum Speed Limit}\label{qsl}
Without loss of generality, we recall the three-level interaction Hamiltonian~Eq.\ref{hamit},
\begin{align}
H(t) &= 
\begin{pmatrix}
0 & \Omega_{ge}(t) & \Omega_{gf}(t) \\
\Omega_{ge}(t) & 0 & \Omega_{ef}(t) \\
\Omega_{gf}(t) & \Omega_{ef}(t) & 0
\end{pmatrix}, \\
\end{align}
with the instantaneous state denoted as $\psi(t) = a(t)|f\rangle + b(t)|e\rangle + c(t)|g\rangle$, initial state denoted as $|\psi\rangle = |g\rangle $ and final state denoted as $|\phi\rangle = |f\rangle.$ To derive a bound on the evolution speed, we define the projection operator onto the target state,
\begin{align}
\Pi_f = |f\rangle \langle f|.
\end{align}

In the Heisenberg picture, the time derivative of the expectation value of \(\Pi_f\) is
\begin{align}
\frac{d}{dt} \langle \Pi_f \rangle &= i \langle [H(t), \Pi_f] \rangle,
\end{align}
where \(\langle \cdot \rangle = \langle \psi(t)| \cdot |\psi(t)\rangle\). It connects the Hamiltonian directly to the rate of change of population in the target state, without explicitly solving the Schrödinger equation. Evaluating the commutator for our Hamiltonian gives
\begin{align}
[H, \Pi_f] &= \Omega_{gf}(t) |g\rangle\langle f| + \Omega_{ef}(t) |e\rangle \langle f| - \mathrm{H.c.},
\end{align}
Taking the expectation value, we obtain
\begin{align}
\frac{d}{dt} |a(t)|^2 & = \frac{d}{dt}  \langle \psi(t)|f\rangle\langle f|\psi(t)\rangle  \\ \nonumber
&= 2\,\mathrm{Im} \big[ a^*(t) (\Omega_{gf}(t) c(t) + \Omega_{ef}(t) b(t)) \big],
\end{align}
therefor
\begin{align}
\frac{d}{dt} |a(t)| & = \frac{\mathrm{Im} \big[ a^*(t) (\Omega_{gf}(t) c(t) + \Omega_{ef}(t) b(t)) \big]}{|a(t)|}.
\end{align}
Defining the angle \(\theta(t)\) via
\begin{align} 
\sin\theta(t) = |a(t)|, \qquad \cos\theta(t) = \sqrt{|b(t)|^2 + |c(t)|^2},
\end{align}
the time derivative of \(\theta(t)\) is
\begin{align} 
|\dot\theta(t)| &= \frac{\frac{d}{dt}|a(t)|}{\sqrt{1 + |a(t)|^2}} \nonumber\\
& \le \frac{| \mathrm{Im}[ a^*(t)(\Omega_{gf} c(t) + \Omega_{ef} b(t)) ] |}{|a(t)|} \nonumber\\
& \le \frac{|a^*(t)|\,|\Omega_{gf} c(t) + \Omega_{ef} b(t)|}{|a(t)|} \nonumber\\
&= |\Omega_{gf}(t)|\,|c(t)| + |\Omega_{ef}(t)|\,|b(t)|.
\end{align}

Including the $\Omega_{ge}(t)$ term gives the full bound
\begin{align}
|\dot\theta(t)| \le |\Omega_{gf}(t)|\,|c(t)| + |\Omega_{ge}(t)|\,|b(t)| + |\Omega_{ef}(t)|\,|b(t)|.
\end{align}

Under the sum-of-squares constraint,
\begin{align}
\Omega_{ge}^2 + \Omega_{ef}^2 + \Omega_{gf}^2 \le \Omega_{\max}^2,
\end{align}
the Cauchy--Schwarz inequality yields
\begin{align}
|\dot\theta(t)| &\le \sqrt{\Omega_{ef}^2 + \Omega_{gf}^2} \sqrt{|b(t)|^2 + |c(t)|^2} \nonumber\\
&\le \Omega_{\max} \sqrt{|b(t)|^2 + |c(t)|^2} \nonumber\\
& = \Omega_{\max} \cos\theta(t).
\end{align}

Under the sum-of-linear constraint,
\begin{align}
|\Omega_{ge}| + |\Omega_{ef}| + |\Omega_{gf}| \le \Omega_{\max},
\end{align}
Hölder's inequality ($p=1$, $q=\infty$) gives
\begin{align}
|\dot\theta(t)| & \le (|\Omega_{ef}| + |\Omega_{gf}|) \max\{|b(t)|,|c(t)|\} \nonumber\\
& \le (|\Omega_{ef}| + |\Omega_{gf}|) \sqrt{|b(t)|^2 + |c(t)|^2} \\
& = \Omega_{\max} \cos\theta(t).
\end{align}

Integrating from \(\theta(0)=0\) to \(\theta(\tau)=\pi/2\) gives the quantum speed limit
\begin{align}
\tau \ge \frac{\pi}{2\Omega_{\max}}.
\end{align}

The bound is saturated when only the direct $g$--$f$ coupling is present,
\begin{align}
\Omega_{ge}(t) = \Omega_{ef}(t) = 0, \quad \Omega_{gf}(t) = \Omega_{\max},
\end{align}
so that the amplitudes reduce to
\begin{align}
c(t) &= \cos(\Omega_{\max} t), &
a(t) &= -i \sin(\Omega_{\max} t), &
b(t) &= 0,
\end{align}
i.e., the evolution remains confined to the $g$--$f$ subspace. In this case, the expression reduces to
\begin{align}
\Delta E = \sqrt{\langle \psi(t)| H^2 | \psi(t) \rangle - E(t)^2} = \Omega_{\max},
\end{align}
so that all the inequalities discussed above are saturated as equalities. This corresponds to the Mandelstam–Tamm form of the QSL:
\begin{align}
    T\{|g\rangle,|f\rangle\}  = \frac{\arccos{|\langle g|f\rangle}|}{\Delta E}
\end{align}
\clearpage
\bibliography{Quantum_battery}
\end{document}